\documentclass[EPiC]{easychair}

\usepackage{doc}

\title{3D Reconstruction from Arthroscopic Images using NeRF: a preliminary in-silico study}

\author{
Hermine Kitio Tsamo\inst{1,2}%
\and 
    Agathe Yvinou\inst{1,4}
\and
    Daniel Pizarro\inst{5}
\and
    Samad Barri Khojasteh\inst{5}
\and
    Alexandre Tronchot \inst{6}
\and 
    Antoine Ferreira \inst{7}
\and
   Eric Stindel\inst{1,4}
\and
   Aziliz Guezou-Philippe\inst{1,3}
\and
    Guillaume Dardenne\inst{1}%
}

\institute{
  LaTIM - UMR 1101 INSERM, Brest, France\\
\and
   University of Western Brittany, Brest, France \\
\and
    IMT Atlantique, Brest, France\\
\and 
    Brest University Hospital, Brest, France \\
\and
   Department of Electronics, Universidad de Alcala, Alcala de Henares, Madrid, Spain\\
\and 
    INSA Centre Val de Loire, Laboratoire PRISME, Bourges, France.
\and
    Orthopaedics and Trauma Department, Rennes University Hospital, Rennes, France
 }

\authorrunning{Hermine KITIO TSAMO et al}

\titlerunning{3D Reconstruction from Arthroscopic Images using NeRF}

\begin{document}

\maketitle

\begin{abstract}
In knee arthroscopy surgery, accurate registration between preoperative and intraoperative anatomy is a critical step for patient-specific navigation. Achieving an accurate registration requires a reliable 3D reconstruction of the joint during surgery. Preoperative 3D models can be obtained from patient imaging through segmentation and reconstruction, but generating an intraoperative 3D representation remains particularly challenging. Arthroscopic imaging suffers from a limited field of view, low surface texture, and strong specular reflections, which make conventional feature-based 3D reconstruction methods unreliable.
In this work, we investigate the application of MIS-NeRF (Minimally-Invasive Surgery Neural Radiance Fields) for reconstructing intraoperative knee 3D models from monocular arthroscopic images. 

The approach is evaluated on six simulated arthroscopic acquisitions representing six patient-specific knee 3D models. Both qualitative and quantitative results are presented to assess the reconstruction quality.

The reconstructed knee 3D models were evaluated through their rendered images, achieving  PSNR (Peak Signal-to-Noise Ratio) of $31.88 \pm 2.82$, SSIM (Structural Similarity Index) of $0.98 \pm 0.004$ and LPIPS (Learned Perceptual Image Patch Similarity) of $0.017 \pm 0.006$.

These preliminary results suggest the feasibility of NeRF-based reconstruction in the challenging context of arthroscopy and may represent a promising step toward accurate in-silico preoperative-to-intraoperative 3D registration for computer-assisted orthopedic surgery. Further, validation on real arthroscopic data will be necessary to assess clinical applicability. 
\end{abstract}



%
%

\section{Introduction}
\label{sect:introduction}

Knee arthroscopy provides intra-articular visualization through a monocular camera and is routinely used in minimally invasive orthopedic surgery. During surgery, the acquired 2D arthroscopic images are essential for surgeons to visualize anatomical structures and treat cartilage lesions. Preoperative CT or MRI imaging is primarily used for diagnosis and patient-specific characterization. To effectively transfer this diagnostic information to the intraoperative setting and support patient-specific treatment, accurate registration between intraoperative arthroscopic images and preoperative 3D knee models derived from CT or MRI is required. Such registration relies on reconstructing a consistent 3D representation of the joint from arthroscopic images, a step that remains particularly challenging due to the limited field of view, partial surface visibility, and the absence of explicit depth information inherent to monocular arthroscopy.

Several approaches have been proposed to reconstruct 3D geometry from monocular images. Classical methods based on visual descriptors or SLAM (simultaneous localization and mapping), such as OneSLAM \cite{teufel2024oneslam}, achieve good performance in textured environments, but their accuracy significantly degrades in knee arthroscopy, where surfaces exhibit very little texture. More recent approaches rely on the joint estimation of depth and camera motion using deep learning models, including EndoSLAM \cite{ozyoruk2021endoslam}, Endo-Depth-and-Motion~\cite{recasens2021endo}, and EndoDAC \cite{cui2024endodac}. While these methods have demonstrated strong potential for 3D reconstruction in endoscopy, their performance also decreases in arthroscopy due to the same visual challenges mentioned above.

Neural Radiance Fields (NeRF) \cite{mildenhall2021nerf} have recently emerged as a powerful approach for dense 3D reconstruction from monocular image sequences. The approach is based on learning a neural function that maps each 3D point and viewing direction to a volumetric density and color, enabling both the synthesis of consistent novel views and the inference of scene geometry. This representation is particularly well suited to monocular imaging setups, where explicit depth information is unavailable. Although initially introduced for natural scenes \cite{arshad2024evaluating}, NeRF-based methods have subsequently been adapted to minimally invasive surgery, notably through approaches such as MIS-NeRF \cite{khojasteh2025mis}, designed for laparoscopic images. These adaptations aim to address visual challenges inherent to laparoscopic imaging, such as variable illumination from the laparoscopic light source, partial surface visibility, and strong specular reflections. However, the application of NeRF-based reconstruction to knee arthroscopy remains largely unexplored.

In this work, we investigate the feasability of partial 3D reconstruction of the knee in the intraoperative setting from simulated monocular arthroscopic images, using MIS-NeRF as the reconstruction framework. Our objective is to assess the potential of this approach in the constrained context of knee arthroscopy.

\section{Methods}
\label{sect:Methods}


\subsection{Dataset generation}

Knee arthroscopy scenes were simulated to evaluate MIS-NeRF in a controlled environment. Six preoperative 3D knee models, reconstructed from patient CT arthroscans, were used to generate six simulated arthroscopic acquisitions within the Unity 3D environment \cite{Unity3D}. Each knee model comprised the femur and tibia along with their associated cartilages. For two models, the anterior cruciate ligament (ACL) was also included. A virtual camera, mimicking an arthroscopic setup, was then moved along trajectories reproducing realistic intra-articular motions, resulting in one acquisition per model with an average of approximately 300 frames. For each sequence, 2D RGB images were generated together with the corresponding camera intrinsic matrix and extrinsic parameters, which were used as inputs to the reconstruction model. The use of ground-truth extrinsic parameters reflects the intended clinical workflow, where the arthroscope pose will be provided by a navigation system during image acquisition.

\subsection{3D model reconstruction from 2D frames}

The architecture of MIS-NeRF is inspired by Nerfacto \cite{tancik2023nerfstudio} and includes specific adaptations to improve robustness to illumination variations and limited visual cues in laparoscopic images. Specifically, the method incorporates a dedicated loss term to exclude specular highlights during training.

In the present work, the specular reflection removal strategy was retained, but the weight of its associated loss term was reduced. Due to the bright and low-texture appearance of articular cartilage, pixels corresponding to valid anatomical structures can exhibit intensity values similar to those of specular highlights, making them difficult to distinguish algorithmically. With the original loss weighting, relevant anatomical regions were sometimes suppressed together with the specular reflections. Reducing the weight of this loss mitigated this issue while preserving the benefits of the original MIS-NeRF framework.

\begin{figure}[h] 
    \centering
    \includegraphics[width=1.1\linewidth]{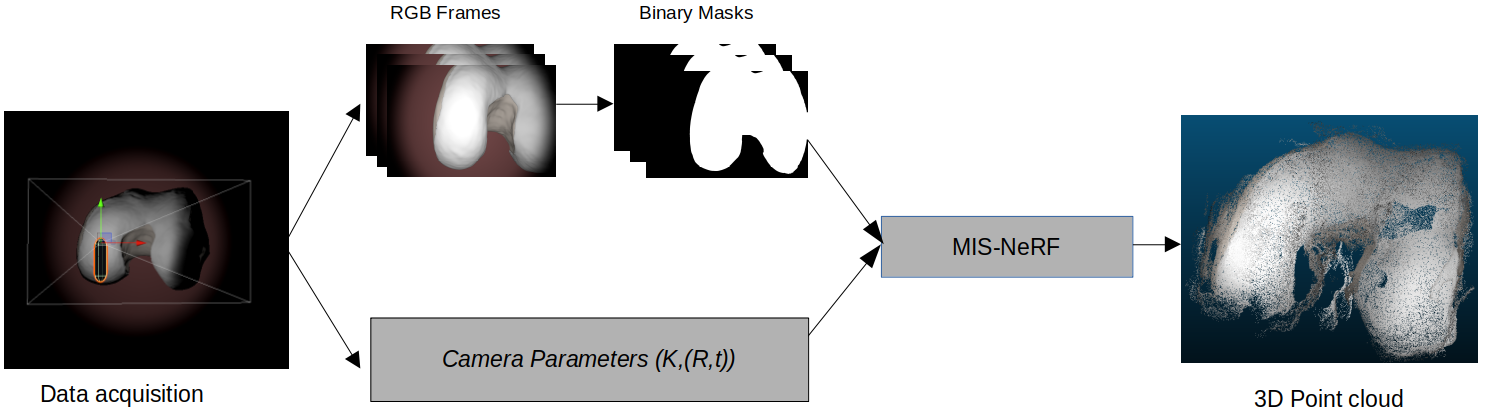}
    \caption{Overview of the proposed workflow. A Unity 3D scene is used to acquire RGB frames and camera parameters. Binary masks are obtained through image pre-processing, including background removal. RGB images, binary masks, and camera parameters (K for intrinsics and (R,t) for extrinsics) are then used as inputs for MIS-NeRF to reconstruct a 3D point cloud.}
    \label{fig:Worflow}
\end{figure}

An overview of the reconstruction pipeline is shown in Figure \ref{fig:Worflow}. At this stage, binary masks are used to restrict the reconstruction to the regions  of interest (femur, tibia, and ACL only). These masks are obtained through a simple color-based segmentation of the background in the RGB images. We do not rely on generic segmentation models such as SAM \cite{kirillov2023segment}, as they are not trained on our simulated arthroscopy images and produce poor results in this context. Unlike the original MIS-NeRF framework, camera intrinsic matrix K and extrinsic parameters (defined by the rotation matrix R and translation vector t) are directly provided by the Unity 3D simulation environment rather than being estimated using structure-from-motion, which is unreliable in our low-texture arthroscopic scenes. As previously mentioned, this setup is consistent with the intended clinical workflow. For each simulated acquisition, 90\% of the frames are used for training and 10\% for evaluation.

\subsection{Reconstruction evaluation}
In NeRF-based reconstruction, performance can be evaluated either at the geometric level by comparing the reconstructed 3D surface with a ground-truth mesh after spatial registration, or at the image level by assessing how accurately the model reproduces observed views. In this work, the registration between the reconstructed geometry and the ground-truth 3D model has not been performed, preventing a direct evaluation of geometric errors.

Consequently, reconstruction quality is assessed using only image-based metrics \cite{tancik2023nerfstudio}. This evaluation compares ground-truth 2D arthroscopic images with images rendered from the trained MIS-NeRF model, providing a reliable preliminary indication of reconstruction fidelity. Three commonly used metrics are employed: PSNR (Peak Signal-to-Noise Ratio), which measures pixel-wise similarity between rendered and reference images; SSIM (Structural Similarity Index), which evaluates structural similarity in terms of luminance, contrast, and local patterns; and LPIPS (Learned Perceptual Image Patch Similarity), which quantifies perceptual differences based on deep feature representations. 

\section{Results}
\label{sect:Results}

\begin{figure*}[h]
\centering
\setlength{\tabcolsep}{4pt}

\begin{tabular}{c c c}
    & \textbf{GT} & \textbf{MIS-NeRF}  \\

\rotatebox{90}{Patient 1} &
\includegraphics[width=0.3\textwidth]{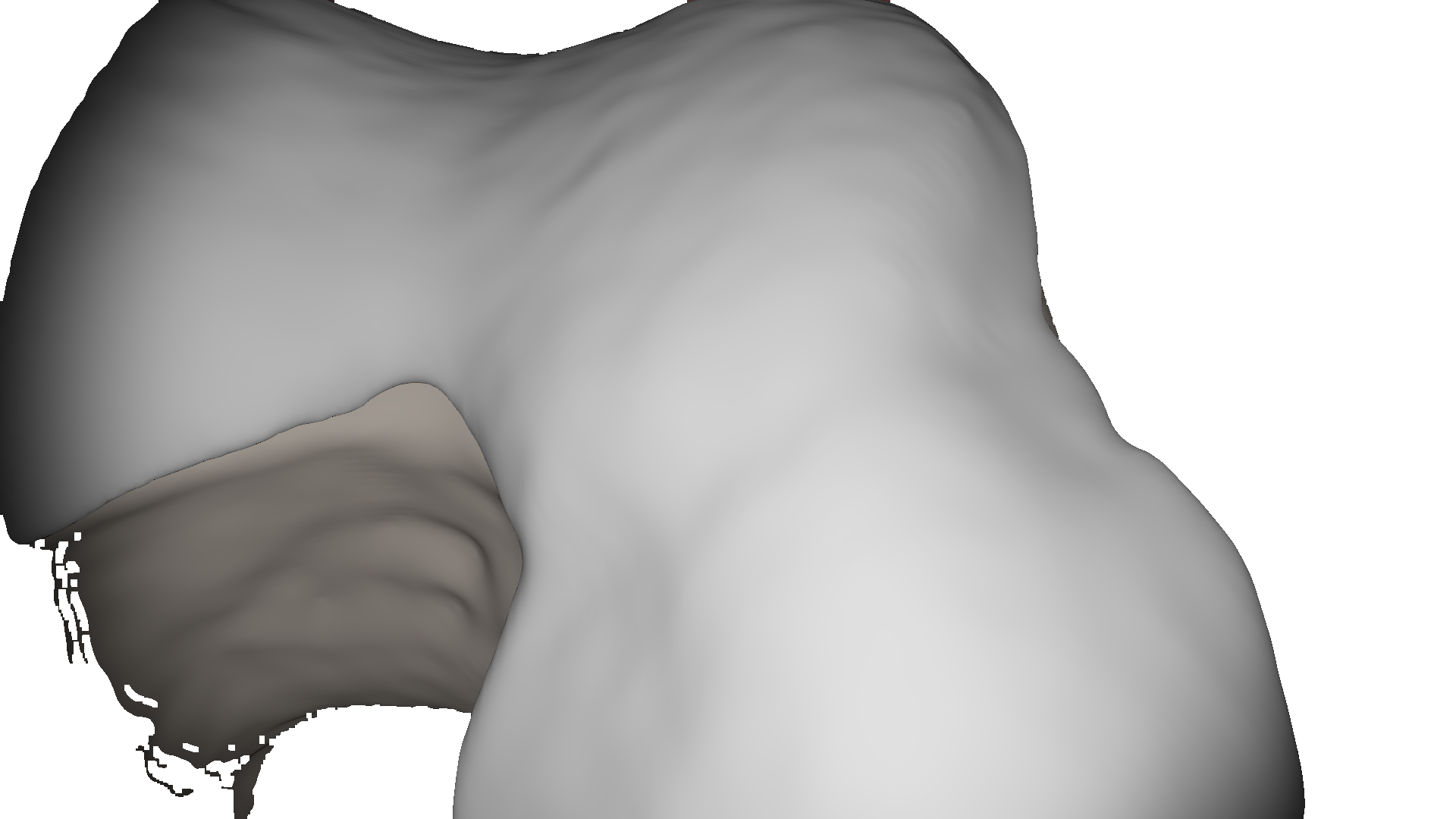} &
\includegraphics[width=0.3\textwidth]{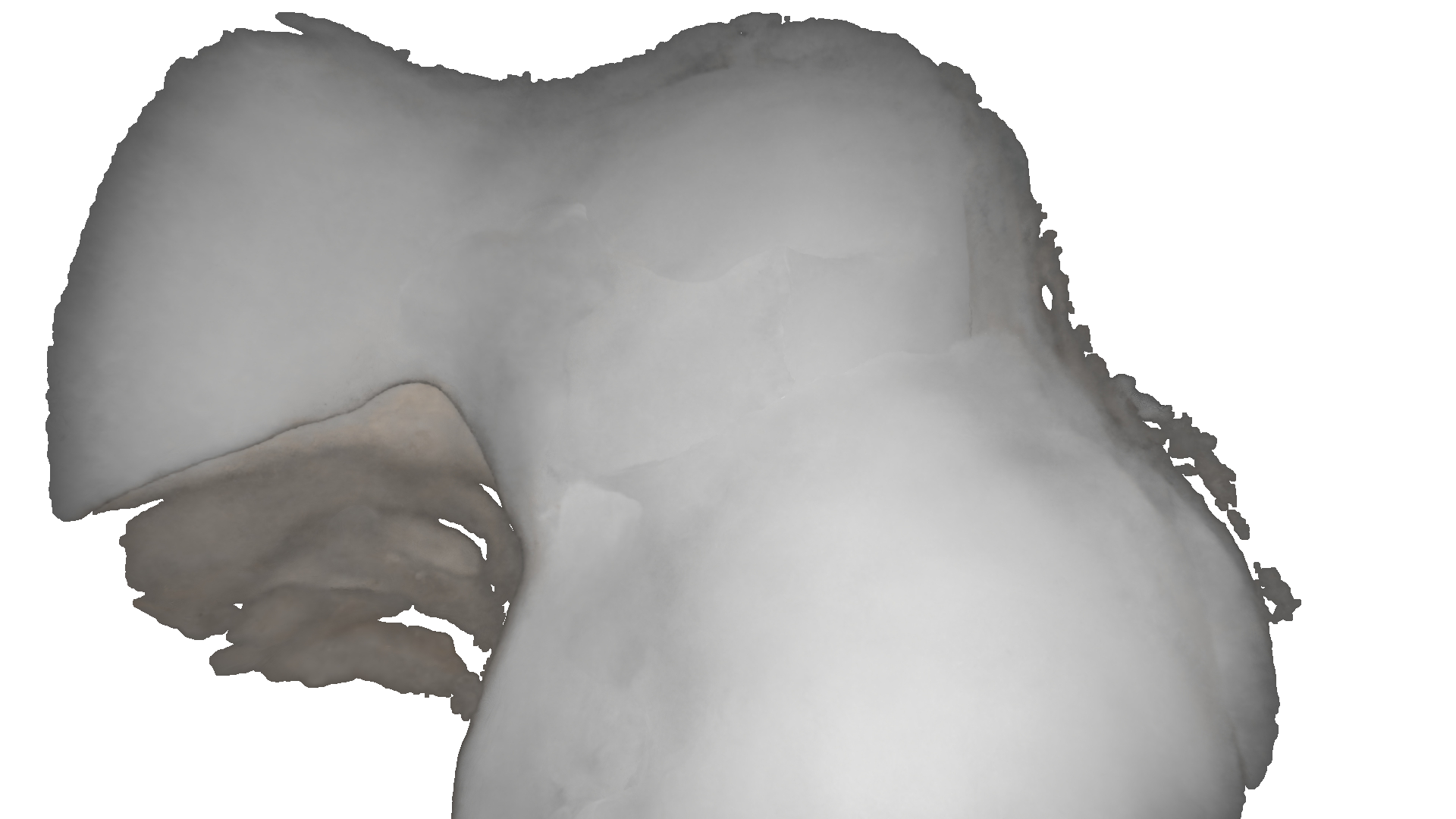} \\

\rotatebox{90}{Patient 2} &
\includegraphics[width=0.3\textwidth]{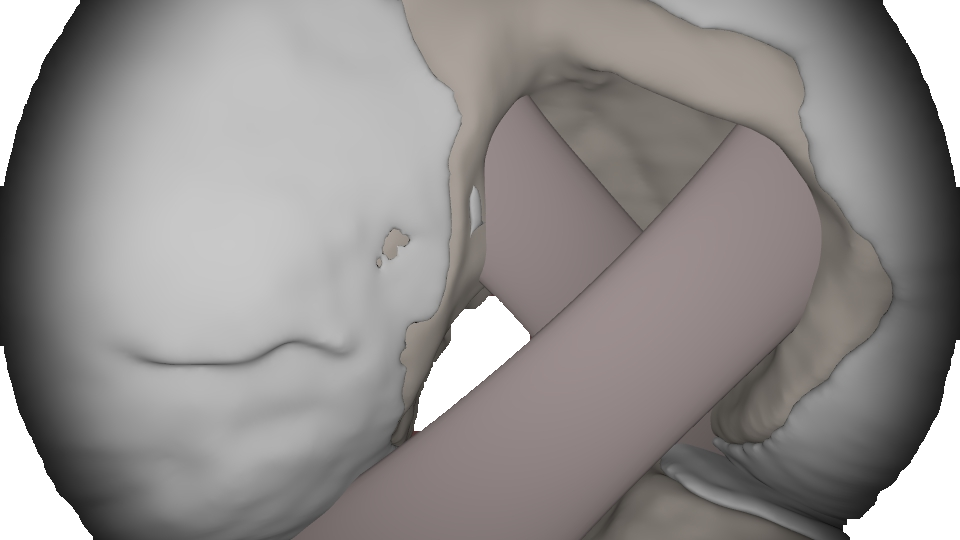} &
\includegraphics[width=0.3\textwidth]{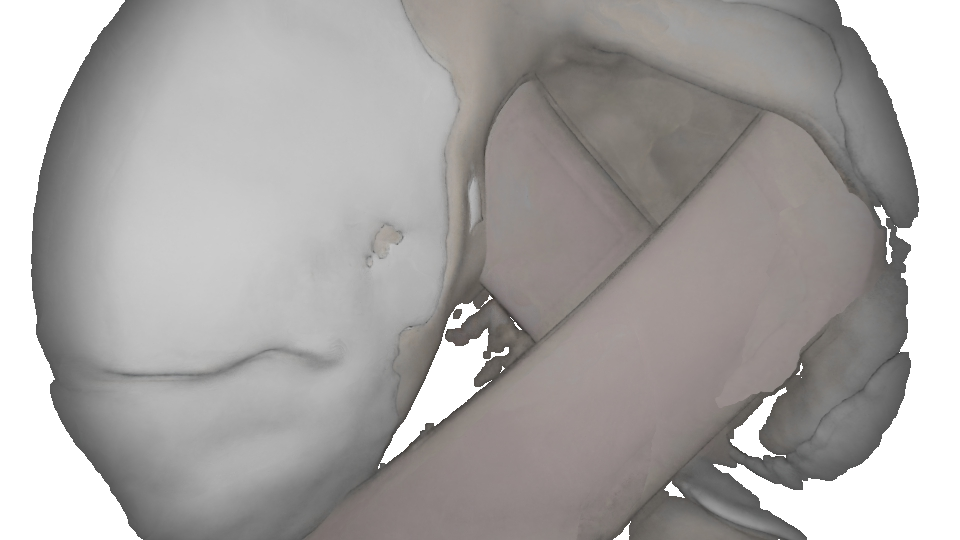} \\

\end{tabular}
\caption{Comparison of ground-truth and MIS-NeRF rendered 2D knee images, with femoral cartilage in white, femur in beige, and cruciate ligaments in pink.}
\label{fig:qualitative_eval}
\end{figure*}

We evaluate the intraoperative 3D reconstructions both qualitatively and quantitatively. Figure \ref{fig:qualitative_eval} presents a visual comparison between the ground-truth 2D images and the rendered images produced by MIS-NeRF for two representative patients' knees.

For the quantitative evaluation, the reconstruction quality was assessed by averaging the results across the six datasets. The model achieved a PSNR of $31.88 \pm 2.82$, an SSIM of $0.98 \pm 0.004$, and an LPIPS of $0.017 \pm 0.006$.


\section{Discussion}
In this work, we investigated the feasibility of applying MIS-NeRF \cite{khojasteh2025mis} to knee arthroscopy, a particularly challenging setting due to low-texture surfaces and constrained intra-articular viewpoints. The presented results demonstrate that MIS-NeRF can reconstruct a coherent 3D representation of the knee from simulated monocular arthroscopic image sequences.

The quantitative results show a PSNR of $31.88 \pm 2.82$, indicating good pixel-level agreement between the rendered and reference images. An SSIM of $0.98 \pm 0.004$ further confirms that structural information is well preserved. In addition, the low LPIPS value of $0.017 \pm 0.006$ indicates a high perceptual similarity between the rendered and reference images, suggesting that the reconstructed appearance remains consistent despite the challenging visual characteristics of arthroscopic scenes. Overall, these results demonstrate that MIS-NeRF generalizes well to simulated knee arthroscopy, achieving reconstruction quality comparable to that reported for liver laparoscopy in the original MIS-NeRF \cite{khojasteh2025mis} study.

Future work will focus on evaluating the geometric accuracy of the reconstructed models for computer-assisted arthroscopy by registering the reconstructed meshes to the corresponding preoperative anatomical models and measuring the Target Registration Error (TRE) using femoral anatomical landmarks. Finally, the proposed framework will be validated on real arthroscopic data to assess its clinical applicability.

\section{Acknowledgments}
\label{sect:acks}
This work was supported by funding from the French government through the National Research Agency (ANR), under grant numbers ANR-23-CPJ1-0131-01 and ANR-23-CE33-0007. The author would like to thank the authors of MIS-NeRF \cite{khojasteh2025mis} for valuable discussions and encouragement. 
This work has been also supported by the Spanish Ministry of Science and Innovation MCIN/AEI/10.13039/501100011033 through project METAMORPH (PID2023-151295OB-I00)

\label{sect:bib}
\bibliographystyle{unsrt}
\bibliography{easychair}


\end{document}